\documentstyle[epsfig,eqsecnum,aps]{revtex}
\begin{document}
\draft
\title{Gamow-Teller strength distributions in $^{76}$Ge and
$^{76}$Se from deformed QRPA}

\author{P. Sarriguren, E. Moya de Guerra}
\address{Instituto de Estructura de la Materia,
Consejo Superior de Investigaciones Cient\'{\i }ficas, \\
Serrano 123, E-28006 Madrid, Spain}
\author{L. Pacearescu, Amand Faessler}
\address{Institute f\"ur Theoretische Physik, Universit\"at T\"ubingen, 
D-72076 T\"ubingen, Germany}
\author{F. \v{S}imkovic}
\address{Department of Nuclear Physics, Comenius University,
SK-842 15, Bratislava, Slovakia}
\author{ A.A. Raduta}
\address{Department of Theoretical Physics and Mathematics, 
Bucharest University, P.O.Box MG11,\\ and Institute of Physics and 
Nuclear Engineering, Bucharest, P.O.Box MG6, Romania}
\date{\today}
\maketitle


\begin{abstract}

We study Gamow-Teller strength distributions of $^{76}$Ge and $^{76}$Se
within a deformed QRPA formalism, which includes residual spin-isospin
forces in the particle-hole and particle-particle channels. We consider 
two different methods to construct the quasiparticle basis, a 
selfconsistent approach based on a deformed Hartree-Fock calculation with 
density-dependent Skyrme forces and a more phenomenological approach
based on a deformed Woods-Saxon potential. Both methods contain pairing 
correlations in the BCS approach. We discuss the sensitivity of 
Gamow-Teller strength distributions to the deformed mean field and 
residual interactions. 
\end{abstract}


\section{Introduction}

The nuclear double $\beta$-decay process is widely considered \cite{2breview} 
as one of the most important sources of information about fundamental issues, 
such as lepton number nonconservation and massive neutrinos, that can be used 
to test the Standard Model.

Theoretically, a condition to obtain reliable estimates for the limits of
the double $\beta$-decay half-lives is that the nuclear structure involved in
the process through the nuclear matrix elements can be calculated correctly.
The proton-neutron quasiparticle random phase approximation (pnQRPA or QRPA 
in short) is one of the most reliable and extended microscopic approximations 
for calculating the correlated wave functions involved in $\beta$ and double 
$\beta$ decay processes. The method was first studied in ref. \cite{halb} to 
describe the $\beta $ strength. It was developed on spherical single-particle 
wave functions and energies with pairing and residual interactions.

The QRPA method was also successfully applied to the description of double 
$\beta$-decay \cite{2bqrpa} after the inclusion of a particle-particle ($pp$) 
residual interaction, in addition to the particle-hole ($ph$) usual channel.
Many more extensions of the QRPA method have been proposed in the literature, 
see ref. \cite{extension} and references therein.

An extension of the pnQRPA method to deal with deformed nuclei was done in 
ref. \cite{kru}, where a Nilsson potential was used to generate single 
particle orbitals. Subsequent extensions including Woods-Saxon type potentials 
\cite{moll}, residual interactions in the particle-particle channel 
\cite{homma}, selfconsistent deformed Hartree-Fock mean fields with consistent 
residual interactions \cite{sarr} and selfconsistent approaches in spherical 
neutron-rich nuclei \cite{doba}, can also be found in the literature.
Nevertheless, the effect of deformation on the double $\beta$-decay processes 
has not been sufficiently studied \cite{raduta,larisa}. 

In ref. \cite{sarr}, ground state and $\beta $-decay properties of exotic 
nuclei were studied on the basis of a deformed selfconsistent HF+BCS+QRPA 
calculation with density dependent effective interactions of Skyrme type.
This is a well founded approach that has been very successful in the 
description of spherical and deformed nuclei within the valley of stability 
\cite{flocard}. In this work we extend those calculations to the study of the 
dependence on deformation of the single $\beta$ branches that build up the 
double $\beta$ process. We focus on the example of the double $\beta$-decay 
of $^{76}$Ge and study $\beta^-$ Gamow-Teller (GT) transitions to the 
intermediate nucleus as well as the $\beta^+$ Gamow-Teller transitions of 
the daughter nucleus $^{76}$Se to the same intermediate nucleus. We discuss 
the similarities and differences of using different single particle mean 
fields of Woods-Saxon (WS) and Hartree-Fock (HF) types. 

In sect. II, we present a brief summary containing the basic points in our
theoretical description. Section III contains the results obtained for the 
bulk properties of $^{76}$Ge and $^{76}$Se and a comparison of our results 
to the experimental available information. In sect. IV we present our results 
for the GT strength distributions and discuss their dependence on the deformed 
mean field and residual interactions. The conclusions are given in sect. V.

\section{Theoretical approach}

In this Section we describe the QRPA formalism used in this work, which is 
based on two different assumptions for the deformed mean field, a Woods-Saxon 
potential and a selfconsistent mean field obtained from a Hartree-Fock 
procedure with Skyrme forces.

In the first approach we use a deformed WS potential with axial symmetry to 
generate single particle energies and wave functions. The parameters of this
potential are taken from the work of Tanaka {\it et al.} \cite{tanaka}. This 
parametrization was proposed originally for spherical nuclei ranging from 
$^{16}$O to $^{208}$Pb, but the derived isospin dependence of the parameters 
allows an extension to deformed nuclei as well. Previous QRPA calculations 
have shown that this parametrization provides realistic levels also for 
deformed nuclei and good results on $M1$ excitations were obtained \cite{m1} 
for nuclei in various mass regions as well.

In these calculations, the quadrupole deformation of the WS potential 
$\beta_2$ is usually determined by fitting the microscopically calculated 
ground state quadrupole moment to the corresponding experimental value. 
The hexadecapole deformation $\beta_4$ is expected to be small for these 
nuclei and we assume it is equal to zero. 

On the other hand, we also perform selfconsistent microscopic calculations 
based on a deformed HF method with density-dependent Skyrme interactions.
We consider in this paper the force Sk3 \cite{beiner} and the force SG2 
\cite{giai} that has been successfully tested against spin and isospin 
excitations in spherical \cite{giai} and deformed nuclei \cite{sarr,sarrnoj}.
For the solution of the HF equations we follow the McMaster procedure that
is based in the formalism developed in ref. \cite{vautherin} as described 
in ref. \cite{vallieres}. Time reversal and axial symmetry are also assumed 
here.

In both schemes, WS and HF, the single-particle wave functions are expanded 
in terms of the eigenstates of an axially symmetric harmonic oscillator in 
cylindrical coordinates, which are written in terms of Laguerre and Hermite 
polynomials. The single-particle states $\left| i\right\rangle $ and their 
time reversed $\left| \bar{i}\right\rangle $ are characterized by the 
eigenvalues $\Omega $ of $J_{z}$, parity $\pi _i$, and energy $\epsilon _{i}$

\begin{equation}
\left| i\right\rangle =\sum_{N}\frac{\left( -1\right) ^{N}+ \pi _{i}}{2}
\sum_{n_{r},n_{z},\Lambda \geq 0,\Sigma } C_{Nn_{r}n_{z}\Lambda 
\Sigma }^{i}\left| Nn_{r}n_{z}\Lambda \Sigma \right\rangle \;  \label{sp_i}
\end{equation}
with $\Omega _{i}=\Lambda +\Sigma \geq \frac{1}{2}$, and

\begin{equation}
\left| \bar{i}\right\rangle =\sum_{N}\frac{\left( -1 \right) ^{N}+
\pi _{i}}{2} \sum_{n_{r},n_{z},\Lambda \geq 0, \Sigma }C_{Nn_{r}n_{z}
\Lambda \Sigma }^{i}\left( -1 \right) ^{\frac{1}{2}-\Sigma }\left| 
Nn_{r}n_{z}-\Lambda -\Sigma \right\rangle  \label{sp_ibar}
\end{equation}
with $\Omega _{\bar{i}}=-\Omega _{i}=-\Lambda -\Sigma \leq -\frac{1}{2}$.
For each $N$ the sum over $n_{r},n_{z},\Lambda \geq 0$ is extended to 
the quantum numbers satisfying $2n_{r}+n_{z}+\Lambda =N.$ The sum over 
$N$ goes from $N=0$ to $N=10$ in our calculations.

Pairing correlations between like nucleons are included in both cases 
in the BCS approximation with fixed gap parameters for protons 
$\Delta _{\pi},$ and neutrons $\Delta _{\nu}$.

The number equation in the neutron sector reads

\begin{equation}
2\sum_{i}v_{i}^{2}=N  \label{numeq}
\end{equation}
where $v_{i}^{2}$ are the occupation probabilities

\begin{equation}
v_{i}^{2}=\frac{1}{2}\left[ 1-\frac{\epsilon _{i}- \lambda _{\nu}}{E_{i}}
\right] \;;\;\;u_{i}^{2}=1-v_{i}^{2}  \label{occu}
\end{equation}
in terms of the quasiparticle energies

\begin{equation}
E_{i}=\sqrt{\left( \epsilon _{i}-\lambda _{\nu} \right) ^{2}+
\Delta _{\nu}^{2}}
\label{qpener}
\end{equation}
These equations are solved iteratively for the WS and HF single-particle 
energies to determine the Fermi level $\lambda _{\nu}$ and the occupation 
probabilities. Similar equations are used to determine the Fermi level and 
occupation probabilities for protons by changing $N$ into $Z$, 
$\Delta _{\nu}$ into $\Delta _{\pi}$, and $\lambda_{\nu} $ into 
$\lambda _{\pi}$.

The fixed gap parameters are determined phenomenologically from the 
odd-even mass differences through a symmetric five term formula involving 
the experimental binding energies \cite{audi}:

\begin{eqnarray}
\Delta _{\nu} &=&\frac{1}{8}\left[ B\left( N-2,Z\right) -4B\left( 
N-1,Z\right) +6B\left( N,Z\right) \right.  \nonumber \\
&&\left. -4B\left( N+1,Z\right) +B\left( N+2,Z\right) \right]  
\label{gaps}
\end{eqnarray}
A similar expression is found for the proton gap $\Delta _{\pi}$ by 
changing $N $ by $Z$ and vice versa. For $^{76}$Ge we obtain 
$\Delta _{\nu}= 1.54$ MeV, $\Delta_{\pi}=1.56 $ MeV and for $^{76}$Se we 
obtain $\Delta _{\nu}= 1.71$ MeV and $\Delta_{\pi}=1.75 $ MeV.

Therefore, at the quasiparticle mean field level, we can observe several
differences with respect to the treatment of the mean field in terms of 
HF or WS potentials. The most important is that the quadrupole deformation 
of the ground state is determined selfconsistently in HF and no explicit 
input parameter is needed. Other differences come from the structure of 
the two-body density-dependent Skyrme force that contains terms absent in 
the WS potential, such as a spin-spin interaction in the selfconsistent 
mean field through the spin exchange operators of the Skyrme force.

Now, we add to the mean field a spin-isospin residual interaction, which 
is expected to be the most important residual interaction to describe GT 
transitions. This interaction contains two parts. A particle-hole part, 
which is responsible for the position and structure of the GT resonance 
\cite{homma,sarr} and a particle-particle part, which is a neutron-proton 
pairing force in the $J^\pi=1^+$ coupling channel.

\begin{equation}
V^{ph}_{GT} = 2\chi ^{ph}_{GT} \sum_{K=0,\pm 1} (-1)^K \beta ^+_K 
\beta ^-_{-K}, \qquad 
\beta ^+_K = \sum_{\pi\nu } \left\langle \nu | \sigma _K |
\pi \right\rangle a^+_\nu a_\pi \, ;
\end{equation}

\begin{equation}
V^{pp}_{GT} = -2\kappa ^{pp}_{GT} \sum_K (-1)^K P ^+_K P_{-K}, \qquad 
P ^+_K = \sum_{\pi\nu} \left\langle \pi \left| \left( \sigma_K\right)^+ 
\right|\nu \right\rangle  a^+_\nu a^+_{\bar{\pi}} \, .
\end{equation}

The two forces $ph$ and $pp$ are defined with a positive and a negative 
sign, respectively, according to their repulsive and attractive character, 
so that the coupling strengths $\chi$ and $\kappa $ take positive values.

The particle-hole residual interaction could in principle be obtained 
consistently from the same Skyrme force used to create the mean field as 
it was done in refs. \cite{sarr} to study exotic nuclei. However, in this
paper we use as a first attempt the coupling strengths from ref. \cite{homma}.
In this reference, the strengths $\chi ^{ph}_{GT}$, and $\kappa ^{pp}_{GT}$
are considered to be smooth functions of the mass number $A$, proportional 
to $A^\mu$. The strength of the $ph$ force is determined by adjusting the 
calculated positions of the GT giant resonances for $^{48}$Ca, $^{90}$Zr 
and $^{208}$Pb. This gives a mass dependence with $\mu =0.7$. The same mass
dependence is assumed for the $pp$ force and the coefficient is determined 
by a fitting procedure to $\beta$-decay half-lives of nuclei with $Z\le 40$. 
The result found in ref. \cite{homma} is $\chi ^{ph}_{GT}= 5.2 \; /A^{0.7}$ 
MeV and $\kappa ^{pp}_{GT}= 0.58\; /A^{0.7}$ MeV. A word of caution is in 
order concerning this parametrization of the residual forces. It serves to 
our purpose of comparing the effects of different deformed mean fields on 
the GT strength distributions, but one should keep in mind that the coupling 
strengths obtained in this way depend in particular, on the model used for 
single particle wave functions and on the set of experimental data considered. 
In ref. \cite{homma} a Nilsson potential was used and the set of experimental 
data did not include the nuclei under study here. Therefore, the coupling 
strengths of ref. \cite{homma} cannot be safely extrapolated and are not 
necessarily the best possible choices. As we shall see in the next sections, 
the strengths from ref. \cite{homma} reproduce well the data when using the 
WS potential, but one needs a somewhat smaller value of $\chi ^{ph}_{GT}$ 
to reproduce the GT resonance with the HF mean field.

The proton-neutron  quasiparticle random phase approximation phonon operator 
for GT excitations in even-even nuclei is written as

\begin{equation}
\Gamma _{\omega _{K}}^{+}=\sum_{\pi\nu}\left[ X_{\pi\nu}^{\omega _{K}}
\alpha _{\nu}^{+}\alpha _{\bar{\pi}}^{+}+Y_{\pi\nu}^{\omega _{K}}
\alpha _{\bar{\nu}} \alpha _{\pi}\right]\, ,  \label{phon}
\end{equation}
where $\alpha ^{+}\left( \alpha \right) $ are quasiparticle creation
(annihilation) operators, $\omega _{K}$ are the RPA excitation energies, and 
$X_{\pi\nu}^{\omega _{K}},Y_{\pi\nu}^{\omega _{K}}$ the forward and backward
amplitudes, respectively. The solution of the QRPA equations can be found 
solving first a dispersion relation \cite{hir}, which is of fourth order in 
the excitation  energies $\omega_K$. 

In the intrinsic frame the GT transition amplitudes connecting the QRPA 
ground state $\left| 0\right\rangle \ \ \left( \Gamma _{\omega _{K}} 
\left| 0 \right\rangle =0 \right)$ to one phonon states 
$\left| \omega _K \right\rangle \ \ \left( \Gamma ^+ _{\omega _{K}} \left| 0
\right\rangle = \left| \omega _K \right\rangle \right)$, are given by
\begin{equation}
\left\langle \omega _K | \sigma _K t^{\pm} | 0 \right\rangle = \mp 
M^{\omega _K}_\pm \, .
\label{intrin}
\end{equation}

The functions $M^{\omega _K}_\pm $ can be found for instance in \cite{hir}.
The basic ingredients in their structure are the spin matrix elements 
connecting neutron and proton states with spin operators
\begin{equation}
\Sigma _{K}^{\nu\pi}=\left\langle \nu\left| \sigma _{K}\right| 
\pi\right\rangle \, ,
\label{sigma}
\end{equation}
which can be written in terms of the coefficients of the expansion in
eqs.(\ref{sp_i})-(\ref{sp_ibar}).

\begin{equation}
\Sigma _{K}^{\nu\pi}=\sum_{Nn_{z}\Lambda \Sigma } C_{Nn_{z}\Lambda
\Sigma +K}^{\nu}C_{Nn_{z}\Lambda \Sigma }^{\pi}\left( 2\Sigma \right) 
\sqrt{1+\left| K\right| }  \label{sig1}
\end{equation}

\begin{equation}
\Sigma _{K=1}^{\nu\bar{\pi}}=\sum_{Nn_{z}}C_{Nn_{z}0 
\frac{1}{2}}^{\nu}C_{Nn_{z}0\frac{1}{2}}^{\pi}\left( -\sqrt{2}\right)  
\label{sig2}
\end{equation}

Once the intrinsic amplitudes are calculated according to eq. (\ref{intrin}),
the GT strength $B(GT)_\pm$ in the laboratory system for a transition 
$I_iK_i (0^+0) \rightarrow I_fK_f (1^+K_f)$ can be obtained as
\begin{equation}
B(GT)_\pm = \frac{g_A^2}{4\pi}\left[ \delta_{K_f,0}\left\langle \phi_{K_f}
\left| \sigma_0t^\pm \right| \phi_0 \right\rangle ^2 +2\delta_{K_f,1}
\left\langle \phi_{K_f} \left| \sigma_1t^\pm \right| \phi_0 \right\rangle ^2
\right],
\label{bgt}
\end{equation}
where we have used the initial and final states in the laboratory frame
expressed in terms of the intrinsic states $\left| \phi_K \right\rangle $
using the Bohr-Mottelson factorization \cite{bohr}.

In the simple uncorrelated 2qp approximation, neglecting the residual $ph$ 
and $pp$ forces, the functions $M^{\omega _K}_\pm$ reduce to the following 
expressions

\begin{equation}
M^{\omega _K}_+ = u_{\nu} v_{\pi} \Sigma _{K}^{\nu\pi} ;\qquad
M^{\omega _K}_- = v_{\nu} u_{\pi} \Sigma _{K}^{\nu\pi} ,
\label{mspm}
\end{equation}
where the excitation energies are the bare two quasiparticle energies
$\omega_K^{\rm 2qp}=E_\nu +E_\pi$.

The Ikeda sum rule is always fulfilled in our calculations

\begin{equation}
\sum_\omega \left[ \left( M^{\omega }_{-}\right) ^2-
\left( M^{\omega }_{+}\right) ^2 \right] = 3(N-Z) \, .
\label{ikedaeq}
\end{equation}

\section{Bulk properties}

In this Section we present results for the bulk properties of $^{76}$Ge and 
$^{76}$Se obtained from WS and HF descriptions.

First, we analyze the energy surfaces as a function of deformation. In the 
case of WS, this is simply done by varying the quadrupole deformation of 
the potential $\beta_2$, which is an input parameter. In the case of HF, 
we perform constrained calculations \cite{constraint}, minimizing the HF 
energy under the constraint of keeping fixed the nuclear deformation. 

We can see in fig. 1 the total energy plotted versus the microscopically
calculated mass quadrupole moment. The results correspond to HF calculations 
with the forces SG2 (solid line) and Sk3 (dashed line), as well as to 
calculations with the WS potential (dotted line). The origin of the energy 
axis is different in each case but the distance between ticks corresponds 
always to 1 MeV.

We observe that the HF calculation predicts the existence of two energy minima 
close in energy, giving rise to shape isomers in these nuclei, while the WS 
potential originates a single energy minimum, which is in agreement with the 
absolute prolate minimum in the case of $^{76}$Ge and close to the prolate HF 
solution in the case of $^{76}$Se.

We can see in Table 1 the experimental and the microscopically calculated 
charge root mean square radii $r_c$, quadrupole moments $Q_p$, and quadrupole 
deformations $\beta$ ($\beta = \sqrt{\frac{\pi}{5}} \frac{Q_p}{Zr_c^2}$). In 
the case of $^{76}$Se, the calculated values correspond to prolate/oblate 
deformations. The input WS prolate deformation is chosen to be $\beta_2=0.10$ 
in both nuclei $^{76}$Ge and $^{76}$Se. In the oblate case of the nucleus 
$^{76}$Se, the WS deformation chosen is  $\beta_2=-0.20$. With these values we 
guarantee that the intrinsic deformations of the ground state are similar in 
HF and WS and therefore the differences in their predictions will have their 
origin in the structure of mean fields having the same deformation.

The values obtained for the charge radii are in good agreement with the 
experimental values from ref. \cite{radiiexp}, which are also shown in 
Table 1. They are also in good agreement with the results obtained from 
relativistic mean field calculations \cite{ring}:
$r_{c ({\rm rel})} (^{76}{\rm Ge})=4.057$ fm and 
$r_{c ({\rm rel})} (^{76}{\rm Se})= 4.119$ fm.

The charge quadrupole moments quoted in Table 1 have been derived 
microscopically from the deformed potentials as ground state expectations 
of the $Q_{20}$ operator. We can compare again with the results from 
relativistic mean field calculations of ref. \cite{ring}:
$Q_{p ({\rm rel})} (^{76}{\rm Ge})= 111.4$ fm$^2$ and 
$Q_{p ({\rm rel})} (^{76}{\rm Se})= -146.8$ fm$^2$. 
These relativistic results are in perfect accordance with our calculated 
results. They can also be compared with experimental intrinsic quadrupole 
moments from ref. \cite{raghavan}. The empirical intrinsic moments are 
derived from the laboratory moments assuming a well defined deformation.
These values are  shown in Table 1 in the first place:
$Q_{p ({\rm exp})} (^{76}{\rm Ge})=66(21)$ fm$^2$ and 
$Q_{p ({\rm exp})} (^{76}{\rm Se})= 119(25)$ fm$^2$.
Experimental quadrupole moments can also be derived \cite{raman} from the 
experimental values of $B(E2)$ strengths, although in this case the sign 
cannot be extracted.
Assuming that the intrinsic electric quadrupole moments are given by
$Q=\sqrt{16\pi B(E2)/5e^2}$, then
$|Q_{p ({\rm exp})}| (^{76}{\rm Ge})=164(24)$ fm$^2$ and 
$|Q_{p ({\rm exp})}| (^{76}{\rm Se})= 205(24)$ fm$^2$.

\section{Gamow-Teller strength distributions}

In this Section we show and discuss the Gamow-Teller strength distributions
obtained from different choices of the deformed mean fields and residual
interactions.

As a general rule, the following figures showing the GT strength distributions 
are plotted versus the excitation energy of daughter nucleus. The 
distributions of the GT strength have been folded with Breit-Wigner functions
of 1 MeV width to facilitate the comparison among the various calculations, 
so that the original discrete spectrum is transformed into a continuous profile. 
These distributions are given in units of $g_A^2/4\pi$ and one should keep in 
mind that a quenching of the $g_A$ factor, typically 
$g_{A,{\rm eff}}=(0.7-0.8)\ g_{A,{\rm free}}$, is expected on the basis of the 
observed quenching in charge exchange reactions.

First of all, we discuss in figs. 2 and 3, the dependence of the GT strength
distributions on the deformed quasiparticle mean field of $^{76}$Ge and 
$^{76}$Se, respectively. To make the discussion meaningful we show the results 
obtained at the two-quasiparticle level without including the spin-isospin 
residual interactions. In these figures we can see the $B(GT_-)$ and  
$B(GT_+)$ strength distributions in the upper and lower panels, respectively. 
One should notice that the relevant strength distributions for the double 
$\beta$-decay of $^{76}$Ge, as it can be seen schematically in fig. 4, are 
the $B(GT_-)$ distribution of the parent $^{76}$Ge and the $B(GT_+)$ 
distribution of daughter $^{76}$Se, but for completeness we show both 
distributions for each nucleus.  Solid lines in figs. 2 and 3 correspond to 
the results obtained from the Skyrme force SG2 within a HF scheme, dashed 
lines correspond to the results obtained with the WS potential. The 
deformation of the mean fields are as indicated in Table 1, using the prolate 
shape in $^{76}$Se. Pairing correlations are included in HF and WS cases in a 
similar way with the gap parameters for neutrons and protons mentioned 
earlier. Then, the only source of discrepancy between HF and WS comes from 
the different single particle wave functions and energies.

In general, we observe that WS and HF produce a similar structure of three 
peaks in the $B(GT_-)$ profiles of $^{76}$Ge and $^{76}$Se, although the WS 
results are somewhat displaced to lower energies with respect to the HF peaks. 
The strengths contained in the peaks are also comparable. In the case of 
the $B(GT_+)$ distributions, we first observe the different scale, which is 
about one order of magnitude lower than the $B(GT_-)$ scale. This is a 
consequence of the Pauli blocking. We can see from eq. (\ref{mspm}) that 
while the occupation amplitudes $u's$ and $v's$ favor $M_-$ strengths, they 
are very small factors in $M_+$ strengths when connecting similar proton and 
neutron states. The difference between total $B(GT_-)$ and $B(GT_+)$ 
strengths (Ikeda sum rule (\ref{ikedaeq}), which is fulfilled in our 
calculations), is a large number $3(N-Z)=36$ in $^{76}$Ge and $3(N-Z)=24$ 
in $^{76}$Se, reflecting the different magnitude of the $B(GT_-)$ and 
$B(GT_+)$ strengths shown in figs. 2 and 3.

The profiles of the $B(GT_+)$ distributions with WS and HF present some
discrepancies that are amplified by the scale. In particular, it is 
remarkable the large strength produced by WS in the region of high excitation 
energies in $^{76}$Ge that we discuss later in terms of the single particle 
wave functions.

In order to clarify the origin of the various peaks in the strength 
distributions we have added in fig. 2 labels showing some of the leading 
transitions generating the strength. The labels stand for 
$pK^{\pi} - nK^{\prime\pi}$ of the orbitals connected by the spin operator 
in eq. (\ref{sigma}) and a number that identifies the transition. In both 
cases, $B(GT_-)$ and $B(GT_+)$, the same type of transitions are connected 
by the GT operator but the occupation probabilities, weighting the matrix 
elements, enhance or reduce them accordingly. We can see from fig. 2 that 
the structure of the profiles in both WS and HF are generated by the same 
type of GT transitions. 

This can be further illustrated by looking at fig. 5, where we show the 
single particle energies for protons and neutrons  obtained in HF(SG2) 
and WS in $^{76}$Ge. In the left part of the figure corresponding to the 
HF calculation we have plotted the occupation probabilities $v_\nu^2$ and 
$v_\pi^2$ and the Fermi energies $\lambda_\nu$ and $\lambda_\pi$. We can 
also see for completeness the spherical levels labeled by their $\ell_j$ 
values. We have indicated by arrows the most relevant Gamow-Teller 
transitions in the $\beta^-$ and $\beta^+$ directions that are labeled 
by the same numbers used in fig. 2 to identify the peaks. To be more 
precise, we can see in Table 2 the correspondence between these labels 
and the transitions connecting the proton and neutron states using the 
asymptotic quantum number notation $[Nn_z\Lambda ]K^\pi$.

Now, looking at fig. 2, we can understand that the two first peaks in 
$B(GT_-)$ are generated mainly by transitions between neutrons and protons 
dominated by contributions within the $N=3$ shell and that the third peak 
is generated by transitions between neutrons and protons with main 
contributions coming from the $N=4$ shell. The different energies of the 
peaks are due to the different concentration of energy levels in HF and WS. 

In the case of $B(GT_+)$, the strength below 8 MeV is mainly generated by 
transitions within the $N=3$ shells. Beyond 8 MeV the strength, which is 
negligible in HF, is generated by transitions between the proton shell 
$N=2$ and the neutron shell $N=4$ as well as between the proton shell 
$N=3$ and the neutron shell $N=5$, always understood as the main components 
of the wave functions. Then, very deep inside protons $(v_p=1)$ are 
connected with very unoccupied neutron states $(u_n=1)$, giving rise to 
maximum occupation factors. The different behavior in this high energy 
region between HF and WS is therefore due, other factors like deformation 
and occupations being equal, to the structure of the deformed orbitals.

To illustrate the role of the different single particle wave functions in 
the development of the peak structure, we consider in detail the case of 
the last peak observed in the $B(GT_+)$ distribution of the WS potential. 
As we can see it is mainly due to a transition between the proton state 
$[303]$ in the $N=3$ shell with $K^{\pi}=7/2^-$ and the neutron state 
$[523]$ in the $N=5$ shell with $K^{\pi}=5/2^-$. The structure of the 
single particle wave functions, according to the expansion in eq. (\ref{sp_i}), 
of these two states in the cases of HF and WS can be seen in Table 3. With 
these coefficients we can construct the matrix elements in eq. (\ref{sig1}). 
The resulting strength is almost two orders of magnitude in favor of WS, 
which explains the huge discrepancy observed between WS and HF in the higher 
energy domain.

Nevertheless, these discrepancies are smaller in the case of the  $B(GT_+)$ 
of $^{76}$Se, which is the relevant branch for the double $\beta$-decay of 
the parent nucleus $^{76}$Ge.

Figures 6 and 7 contain the strength distributions obtained from QRPA 
calculations for $^{76}$Ge and $^{76}$Se, respectively. The data in Fig. 6 
are from ref. \cite{geexp} and were obtained from charge exchange 
$^{76}$Ge(p,n)$^{76}$As reactions. The thick line in Fig. 6 corresponds
to these data folded by the same procedure used for the theoretical results. 
The data in Fig. 7 and 8 are from ref. \cite{seexp} and were obtained from 
charge exchange $^{76}$Se(n,p)$^{76}$As reactions.

The coupling constants of the $ph$ and $pp$ residual interactions used in 
Figs 6-8 are from ref. \cite{homma} in the case of WS. In our case with $A=76$, 
these parameters are $\chi ^{ph}_{GT}=0.25$ MeV and $\kappa ^{pp}_{GT}=0.027$ 
MeV. In the case of the HF calculations with the Skyrme forces Sk3 and SG2
better agreement with the measured location of the $B(GT_-)$ resonance in 
$^{76}$Ge is obtained with a somewhat smaller value of the $ph$ strength. 
The curves shown in Figs. 6-8 for the HF results have been obtained using  
$\chi ^{ph}_{GT}=0.16$ MeV and the same $\kappa ^{pp}_{GT}=0.027$ MeV.

In Fig. 6 we have used the prolate deformations for $^{76}$Ge given in Table 1.
We can see that WS follows the structure of the experimental $B(GT_-)$ strength
distribution with two peaks at low energies ($E_{ex}=$5 and 8 MeV) and the 
resonance at 11 MeV. The HF calculations produce also a few peaks at low
excitation energies and a resonance between 10 and 13 MeV. We can see that the 
structure of the strength distributions is qualitatively similar for the two 
Skyrme forces and that the difference with the WS curves can be traced back 
to the discrepancies found at the two quasiparticle level.

Fig. 7 contains similar calculations for $^{76}$Se. The coupling strengths of
the residual forces are as indicated for $^{76}$Ge. The results in the HF cases
are obtained with the oblate deformation of $^{76}$Se that produces the
absolute minimum of the energy and agrees better with the experimental quadrupole
moment. In general, comparison with experiment is reasonable and should not be
stressed too much since, as stated in ref. \cite{seexp}, the experimental results, 
especially above 6 MeV, must be considered to be of a qualitative nature only.

The role of the residual interactions on the GT strengths was already studied
in ref. \cite{sarr}, where it was shown that the repulsive $ph$ force 
introduces two types of effects: A shift of the GT strength to higher 
excitation energies with the corresponding displacement of the position of 
the GT resonance and a reduction of the total GT strength. The residual $pp$, 
being an attractive force, shifts the strength to lower excitation energies, 
reducing the total GT strength as well. Also shown in ref. \cite{sarr} was 
the effect of the BCS correlations on the GT strength distribution. The main 
effect of pairing correlations is to create new transitions that are 
forbidden in the absence of such correlations. The main effect of increasing 
the Fermi diffuseness is to smooth out the profile of the GT strength 
distribution, increasing the strength at high energies and decreasing the 
strength at low energies.

The role of deformation was also studied in ref. \cite{sarr}, showing that
the GT strength distributions corresponding to deformed nuclear shapes are 
much more fragmented than the corresponding spherical ones, as it is clear 
because deformation breaks down the degeneracy of the spherical shells.
It was also shown that the crossing of deformed energy levels that depends 
on the magnitude of the quadrupole deformation as well as on the oblate or 
prolate character, may lead to sizable differences between the GT strength 
distributions corresponding to different shapes.

We can see in fig. 8 the GT strength distributions in $^{76}$Se obtained from
spherical, prolate and oblate shapes. They correspond to QRPA calculations 
performed with the HF basis obtained with the force Sk3. In the spherical 
case, the only possible transitions (see fig. 5) are those connecting 
spherical $\ell_j$ partners with $\Delta \ell=0,\; \Delta j=0,1$, in allowed 
approximation. Therefore, there is GT strength only at a few excitation 
energies. The strength we observe in fig. 8 is the result of the folding 
procedure performed at these energies. On the other hand, in the deformed 
cases we can observe a stronger fragmentation, which is the result of all 
possible connections among the deformed states (see fig. 5). Thus, the 
spherical peaks become broader when deformation is present. 

We can see in Table 4 the total GT strengths in $^{76}$Se contained below 
an energy cut of 60 MeV. We show the results obtained for $\beta^+$ and 
$\beta^-$ strengths with oblate, spherical, and prolate shapes. The Ikeda 
sum rule $3(N-Z)=24$, is fulfilled at this energy cut within a $0.3\%$ 
accuracy. We can see from Table 4 that deformation increases both $\beta^+$ 
and $\beta^-$ strengths in a similar amount in order to preserve the Ikeda 
sum rule ($\beta^- - \beta^+$). We also show for comparison the results 
obtained in 2qp approximation. We can see the reduction of the strength 
introduced by the QRPA correlations, which is again similar in absolute 
terms for $\beta^+$ and $\beta^-$ strengths in order to keep the Ikeda sum 
rule conserved in QRPA. Since the $\beta^-$ strength is much larger than 
the $\beta^+$ strength, the relative effect of the QRPA correlations is 
much stronger for  $\beta^+$, where the total strength is reduced by a 50\%.

Comparing the results for $^{76}$Se obtained at different deformations with 
the selfconsistent mean fields (HF with Sk3) in Fig. 8 and Table 4, we see 
that there is a strong dependence on deformation in the strength distributions 
as a function of the energy. However, the total strength does not depend so 
much on deformation. There is an increase of a few percent in going from the 
spherical to the oblate and prolate shapes. The latter observation enters in 
contradiction with SU(3) and shell model calculations by previous authors 
\cite{auerbach} on the dependence on deformation of the GT strengths in 
$^{20}$Ne and $^{44}$Ti. We think that this is due to the much larger and 
richer single particle basis used in the present calculations. In our case 
each single particle state contains mixtures from many harmonic oscillator 
shells (up to $N=10$), while in the above mentioned calculations 
\cite{auerbach}, the single particle basis is restricted to a single harmonic
oscillator major shell (the $sd$ shell in $^{20}$Ne and the $fp$ shell in 
$^{44}$Ti). On the other hand, one may question whether in the deformed cases 
the total strengths calculated here may contain spurious contributions from 
higher angular momentum components in the initial and final nuclear wave 
functions. Since the matrix elements of the transition operator, which is a 
dipole tensor operator, are taken between the states considered in the 
laboratory frame (see eq. (\ref{bgt})), the effect of angular momentum 
projection is to a large extent taken into account. 
We have calculated an upper bound to such contributions using angular 
momentum projection techniques \cite{elvira}. We find that this upper 
bound is less than $1\%$ ($\sim \left\langle J_\perp ^2\right\rangle ^{-2}$, 
with $\left\langle J_\perp ^2\right\rangle=19$ for the oblate shape in 
$^{76}$Se). Thus, exact angular momentum projection would not wash out the 
small increase of the total strength with deformation. 

\section{Concluding remarks}

We have studied the GT strength distributions for the two decay branches 
$\beta^-$ and $\beta^+$ in the double $\beta$-decay of $^{76}$Ge. This has 
been done within a deformed QRPA formalism, which includes $ph$ and $pp$ 
separable residual interactions. The quasiparticle mean field includes
pairing correlations in BCS approximation and it is generated by two 
different methods, a deformed HF approach with Skyrme interactions and a 
phenomenological deformed WS potential. One difference is that with HF and 
Sk3 we get the minimum and stable deformation for $^{76}$Se to be oblate, 
while the prolate minimum is comparable to that obtained with WS and it 
is higher in energy.

We have studied the similarities and differences observed in the GT strength
distributions with these two methods. Among the similarities we can mention
the structure of peaks found in the strength distributions and among the
differences the displacement in the excitation energies found between HF and
WS results. This discrepancy has its origin in the structure of the single 
particle wave functions and  energies generated by the deformed mean fields. 
This also implies that different mean fields require different residual 
interactions to reproduce the experimental GT resonances.

Therefore, in order to obtain reliable GT strength distributions and 
consequently reliable estimates for double $\beta$-decay half-lives, it is 
important to have not only the proper residual interactions but also a good 
deformed single particle basis as a starting point. 
In the case of HF we have seen that standard Skyrme forces, such as SG2 or Sk3,
give a good description of the GT strength distributions provided the proper 
residual interactions are included. Even though the selfconsistent HF approach 
is a more sophisticated type of calculation, the deformed WS potential produces 
comparable results when the parameters of the potential and the residual 
interactions for a given mass region are chosen properly.

There is work in progress to extend these calculations to the double $\beta$-decay 
process studying the dependence on deformation of the half-lives.

\begin{center}
{\Large \bf Acknowledgments} 
\end{center}
This work was supported by Ministerio de Ciencia y Tecnolog\'{\i}a (Spain) 
under contract numbers PB98/0676 and BFM2002-03562 and by International
Graduiertenkolleg GRK683, by the ``Land Baden-Wuerttemberg'' within the 
``Landesforschungsschwerpunkt: Low Energy Neutrinos'', and by the DFG under
436SLK 17/2/98.
\vfill\eject

\newpage
\begin{table}[t]

{\bf Table 1.}
Charge root mean square radii $r_c$ [fm], intrinsic charge quadrupole moments
$Q_p$ [fm$^2$], and quadrupole deformations $\beta$ for $^{76}$Ge and $^{76}$Se
calculated with various assumptions for the mean field.  In the case of 
$^{76}$Se we show theoretical values corresponding to the prolate shape in 
first place and to the oblate shape in second place. Experimental values for 
$r_c$ are from \cite{radiiexp} and for $Q_p$ from \cite{raghavan} the first 
value and from \cite{raman} the second (see text).

\vskip 0.5cm
\begin{center}
\begin{tabular}{llccc}
& & $r_c$ & $Q_p$ & $\beta$ \\ 
\hline \\
$^{76}$Ge & exp. & 4.080 - 4.127 & 66(21) - 164(24) & 0.10 - 0.24 \\ \\
          & Sk3  & 4.130 & 111.0 & 0.161   \\
          & SG2  & 4.083 & 105.9 & 0.157   \\
          & WS   & 3.950 & 110.9 & 0.176   \\
\hline \\
$^{76}$Se & exp. & 4.088 - 4.162 & 119(25) - 205(24) & 0.16 - 0.29 \\ \\
          &      & prol / obl    &   prol / obl   & prol / obl \\
          & Sk3  & 4.170 / 4.180 & 117.5 / -136.0 & 0.158 / -0.181  \\
          & SG2  & 4.113 / 4.143 & 35.2 / -140.6 & 0.049 / -0.191  \\
          & WS   & 3.991 / 4.138 & 81.6 / -141.4 & 0.119 / -0.193  \\
\end{tabular}
\end{center}
\end{table}

\vskip 0.5cm
\begin{table}[t]

{\bf Table 2.}
Correspondence of the labels used in figs. 2 and 5 with the asymptotic 
quantum numbers notation $[Nn_z\Lambda ]K^\pi$.

\begin{center}
\begin{tabular}{ccccc}
& $\beta^-$ & & $\beta^+$ \\ \\
(1) & $\nu [301]1/2^- \rightarrow \pi [301]3/2^-$ &
    & $\pi [303]7/2^- \rightarrow \nu [303]5/2^-$  \\
(2) & $\nu [301]3/2^- \rightarrow \pi [301]1/2^-$ &
    & $\pi [312]5/2^- \rightarrow \nu [303]5/2^-$  \\
(3) & $\nu [303]7/2^- \rightarrow \pi [303]5/2^-$ &
    & $\pi [312]5/2^- \rightarrow \nu [312]3/2^-$  \\
(4) & $\nu [312]5/2^- \rightarrow \pi [312]3/2^-$ &
    & $\pi [202]3/2^+ \rightarrow \nu [413]5/2^+$  \\
(5) & $\nu [420]1/2^+ \rightarrow \pi [440]1/2^+$ &
    & $\pi [330]1/2^- \rightarrow \nu [530]1/2^-$  \\
(6) & & & $\pi [303]7/2^- \rightarrow \nu [523]5/2^-$  \\
\end{tabular}
\end{center}
\end{table}


\begin{table}[t]

{\bf Table 3.} Main coefficients $C^i_{\alpha}$ in the expansion of
eq. (\ref{sp_i}) for the proton state $[303]$ with $K^\pi =7/2^-$ and the
neutron state $[523]$ with $K^\pi =5/2^-$. This is the main contribution 
to the peak at 15 MeV  in the $B(GT_+)$ strength distribution of $^{76}$Ge
with the Woods-Saxon potential. The basis states are labeled by 
$\left| N n_z \Lambda \right\rangle$ quantum numbers. The table contains 
also the contributions from these basis states to the spin matrix elements 
in eqs. (\ref{sigma})-(\ref{sig1}).

\vskip 0.5cm
\begin{center}
\begin{tabular}{llcccccc}
& & $\left| 303\right\rangle$ & $\left| 503\right\rangle$ & 
$\left| 523\right\rangle$ & $\left| 703\right\rangle$ & 
$\left| 723\right\rangle$ & $\left| 903\right\rangle$\\ 
\hline
$7/2^-$ proton & & & & & \\
 & HF(SG2) & -0.9742 & 0.2204 & -0.0061 & 0.0219 & -0.0272 & 0.0122 \\
 & WS      &  0.9876 &-0.1400 & 0.0563 & -0.0233 & 0.0107 & -0.0295 \\
\hline
$5/2^-$ neutron & & & & &  \\
 & HF(SG2) &  0.1369 & 0.5933 & -0.5031 & -0.3928  & 0.2349 & 0.2385 \\
 & WS      & -0.2397 &-0.5173 & 0.5049 &  0.3794 & -0.2596 & -0.2056 \\
\hline
 contribution to $\Sigma _{K}^{\nu\pi}$  & & & & &  \\
 & HF(SG2) & -0.1333 & 0.1308 & 0.0031 & -0.0107 & -0.0064 & 0.0029  \\
 & WS      & -0.2367 & 0.0724 & 0.0284 & -0.0088 & -0.0028 & 0.0061 \\
\end{tabular}
\end{center}
\end{table}
\newpage

\begin{table}[t]

{\bf Table 4.} Total Gamow-Teller strength in $^{76}$Se calculated with 
the force Sk3. Results correspond to $\beta^+$ and $\beta^-$ strengths 
for the oblate, spherical, and prolate shapes calculated in 2qp and QRPA
approximations. All the GT strength contained below an excitation energy
of 60 MeV has been included.  

\vskip 0.5cm
\begin{center}
\begin{tabular}{lcccccccccc}
&&\multicolumn{2}{c}{oblate} && \multicolumn{2}{c}{spherical} && 
\multicolumn{2}{c}{ prolate} \\
&& $\beta^+$ & $\beta^-$ && $\beta^+$ & $\beta^-$ && $\beta^+$ & $\beta^-$\\ 
\hline \\
RPA && 2.420 & 26.331 && 1.846 & 25.765 && 2.599 & 26.524 \\
2qp && 4.387 & 28.298 && 3.816 & 27.736 && 4.971 & 28.892
\end{tabular}
\end{center}
\end{table}

\begin{center}

\begin{figure}[t]
\epsfig{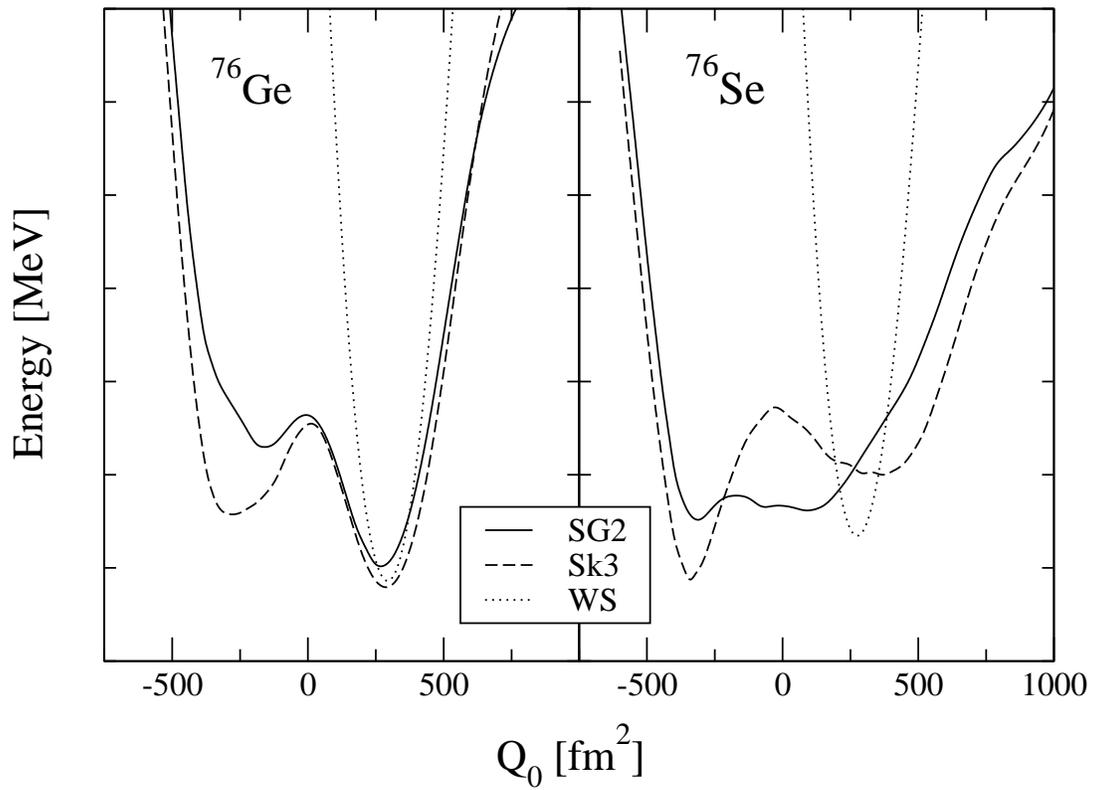}
\vskip 1cm
\caption{Total energy as a function of the mass quadrupole moment
obtained from deformed Hartree-Fock calculations with the Skyrme
forces SG2 (solid line) and Sk3 (dashed lines), and from deformed
Woods-Saxon potentials (dotted line). The origin of the energy axis 
is different in each case but the distance between ticks corresponds 
always to 1 MeV.}
\end{figure}

\begin{figure}[t]
\epsfig{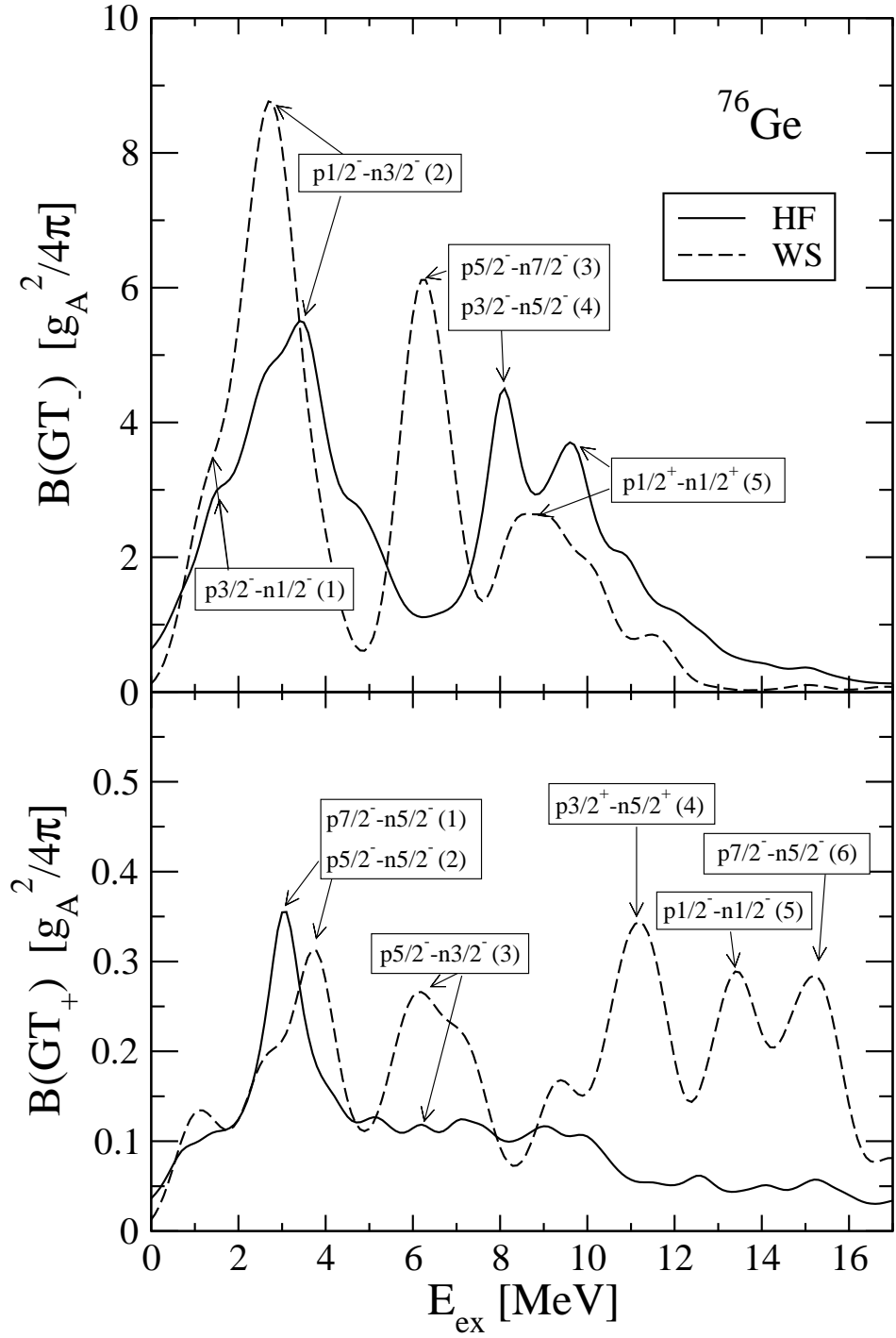}
\vskip 1cm
\caption{Gamow-Teller $B(GT_-)$ and $B(GT_+)$ strength distributions 
$[g_A^2/4\pi]$ in $^{76}$Ge plotted as a function of the excitation energy 
of the daughter nucleus. We compare results of HF(SG2)+BCS 
(solid line) and WS+BCS (dashed line) approximations for the prolate 
minima.}
\end{figure}

\begin{figure}[t]
\epsfig{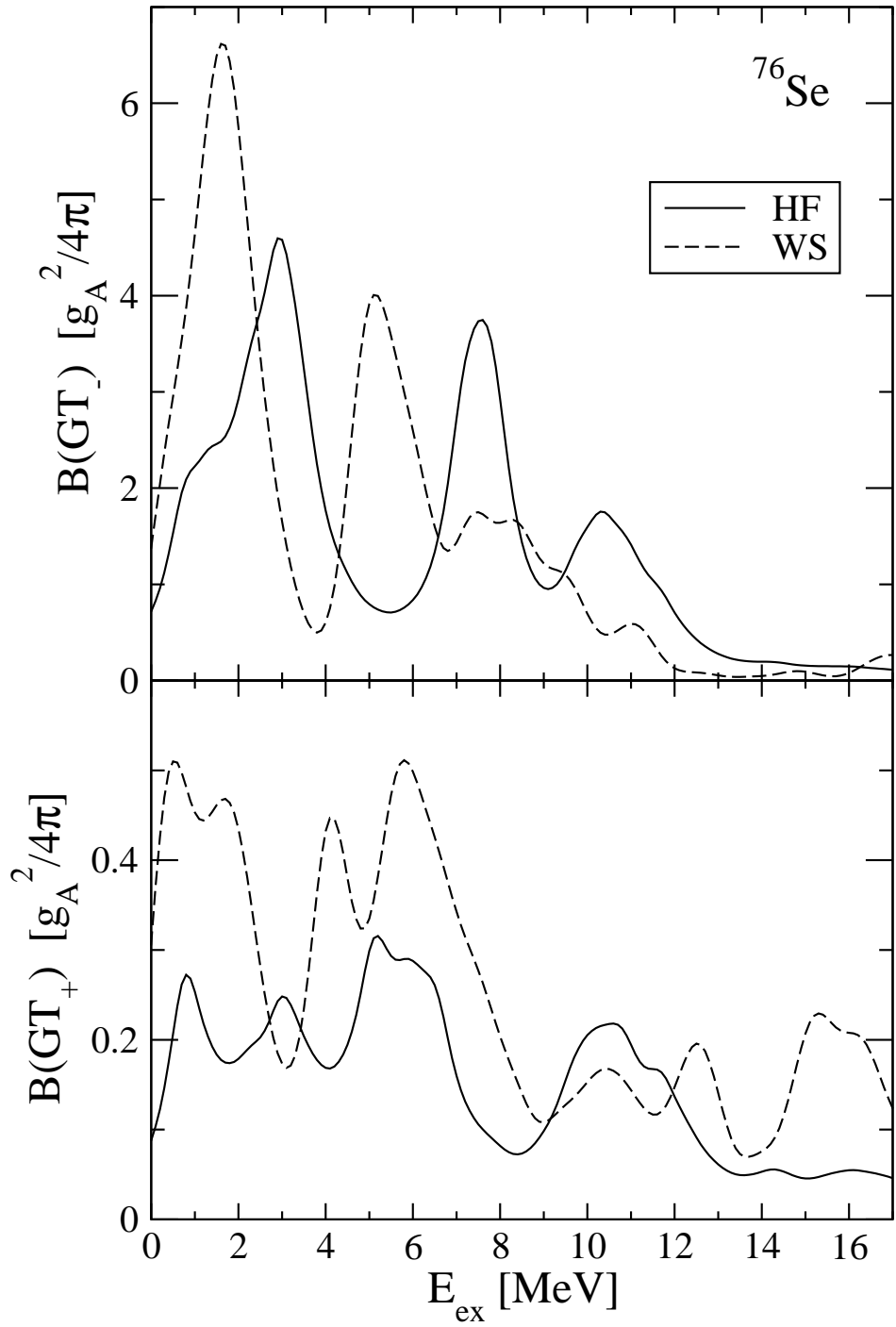}
\vskip 1cm
\caption{Same as in fig. 2 for $^{76}$Se.}
\end{figure}

\begin{figure}[t]
\epsfig{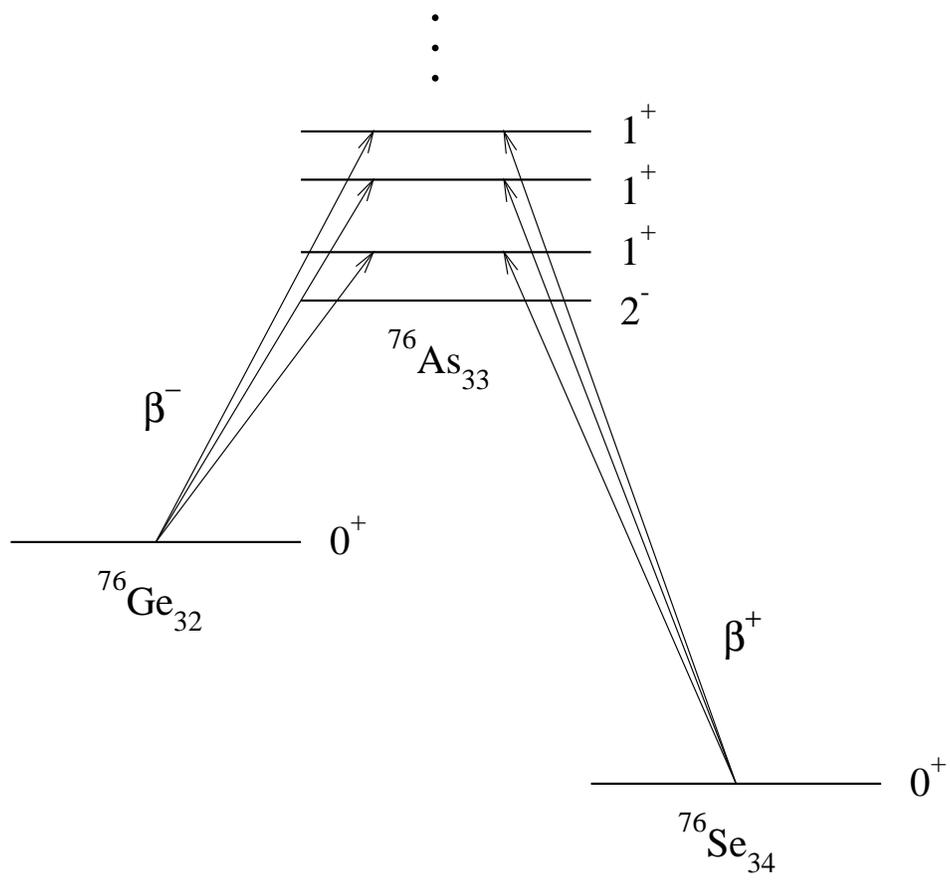}
\vskip 1cm
\caption{Schematic picture of the GT $\beta^- (\beta^+)$ decay of $^{76}$Ge 
($^{76}$Se) into $^{76}$As.}
\end{figure}

\newpage

\begin{figure}[t]
\epsfig{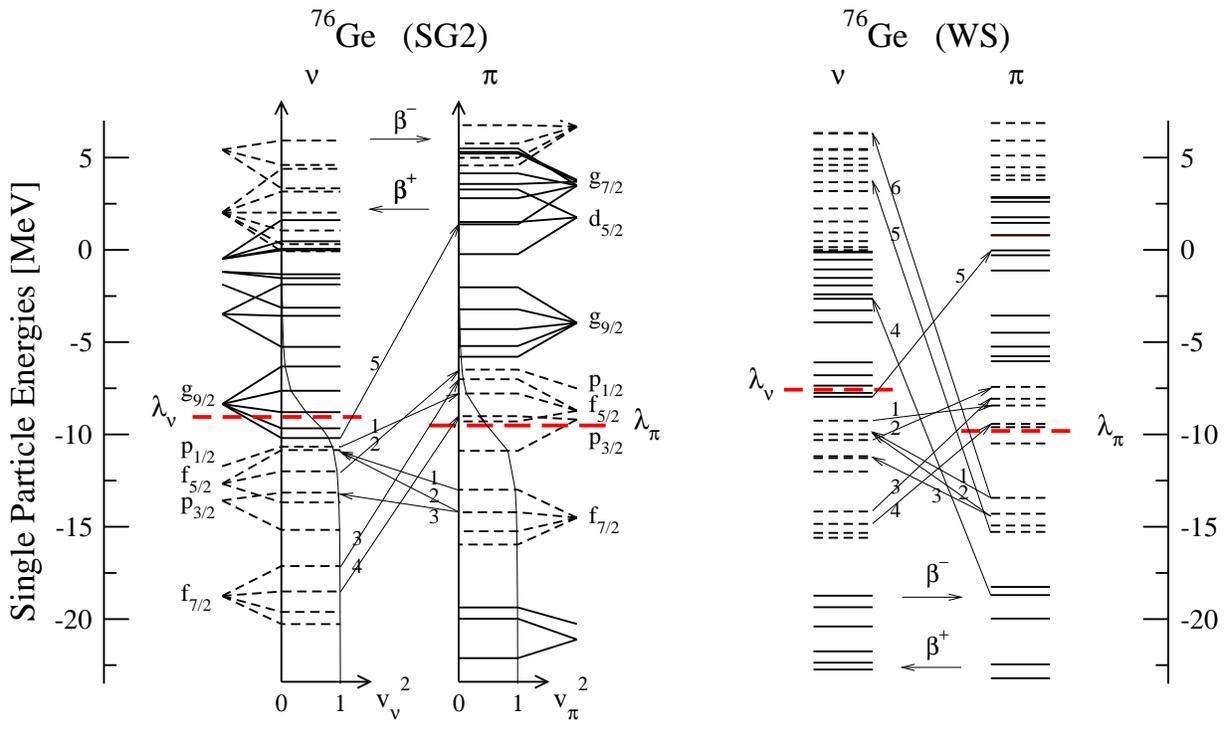}
\vskip 1cm
\caption{Hartree-Fock and Woods-Saxon single particle energies for protons and 
neutrons in $^{76}$Ge.}
\end{figure}
\newpage

\begin{figure}[t]
\epsfig{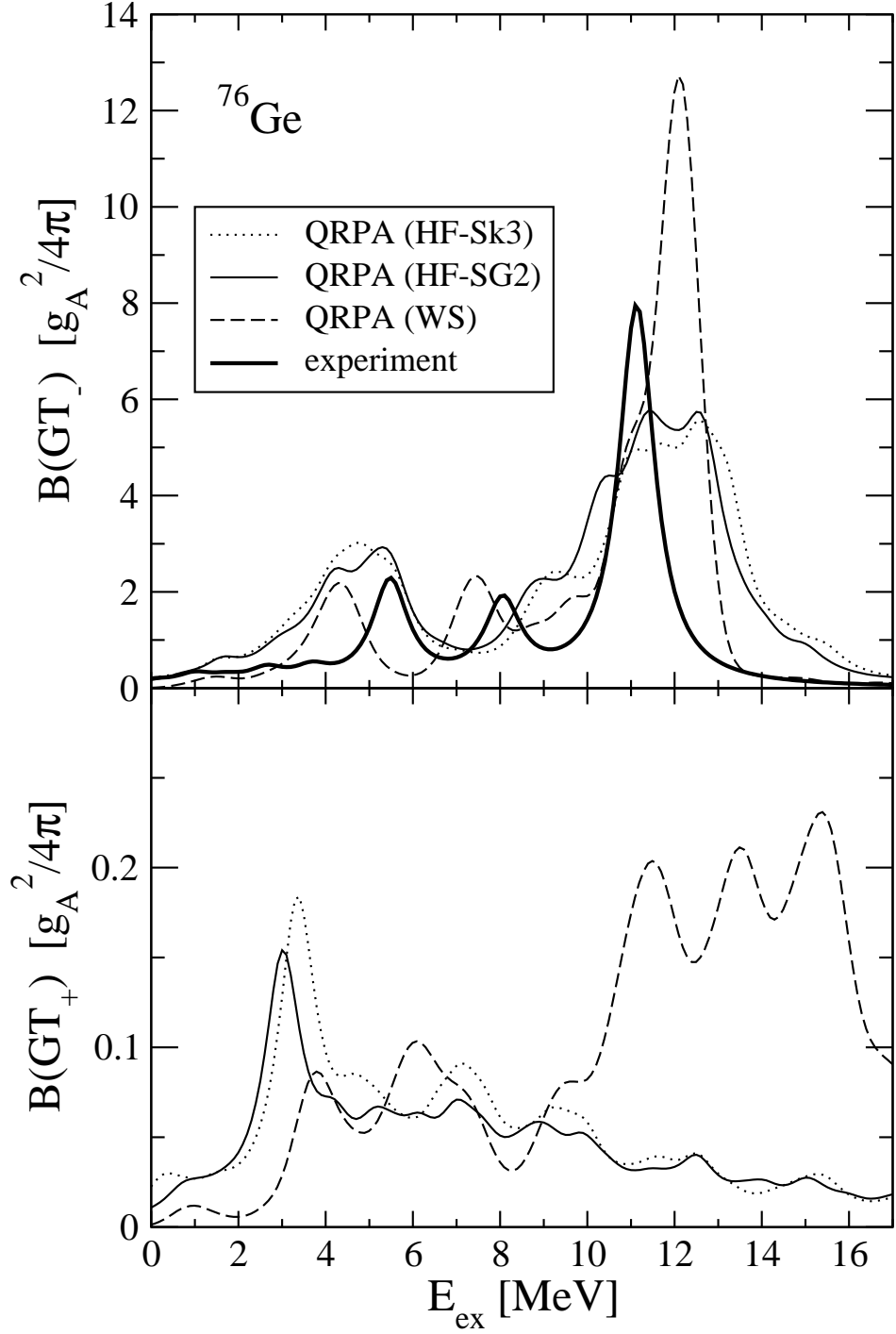}
\vskip 1cm
\caption{Gamow-Teller $B(GT_-)$ and $B(GT_+)$ strength distributions 
$[g_A^2/4\pi]$ in $^{76}$Ge plotted as a function of the excitation energy 
of the daughter nucleus. We compare QRPA results of HF(SG2) (solid lines),
HF(Sk3) (dotted lines) and WS (dashed lines). Deformations and coupling strengths 
of the residual interactions are given in the text.
Experimental data (thick solid lines) are from [28].}
\end{figure}

\begin{figure}[t]
\epsfig{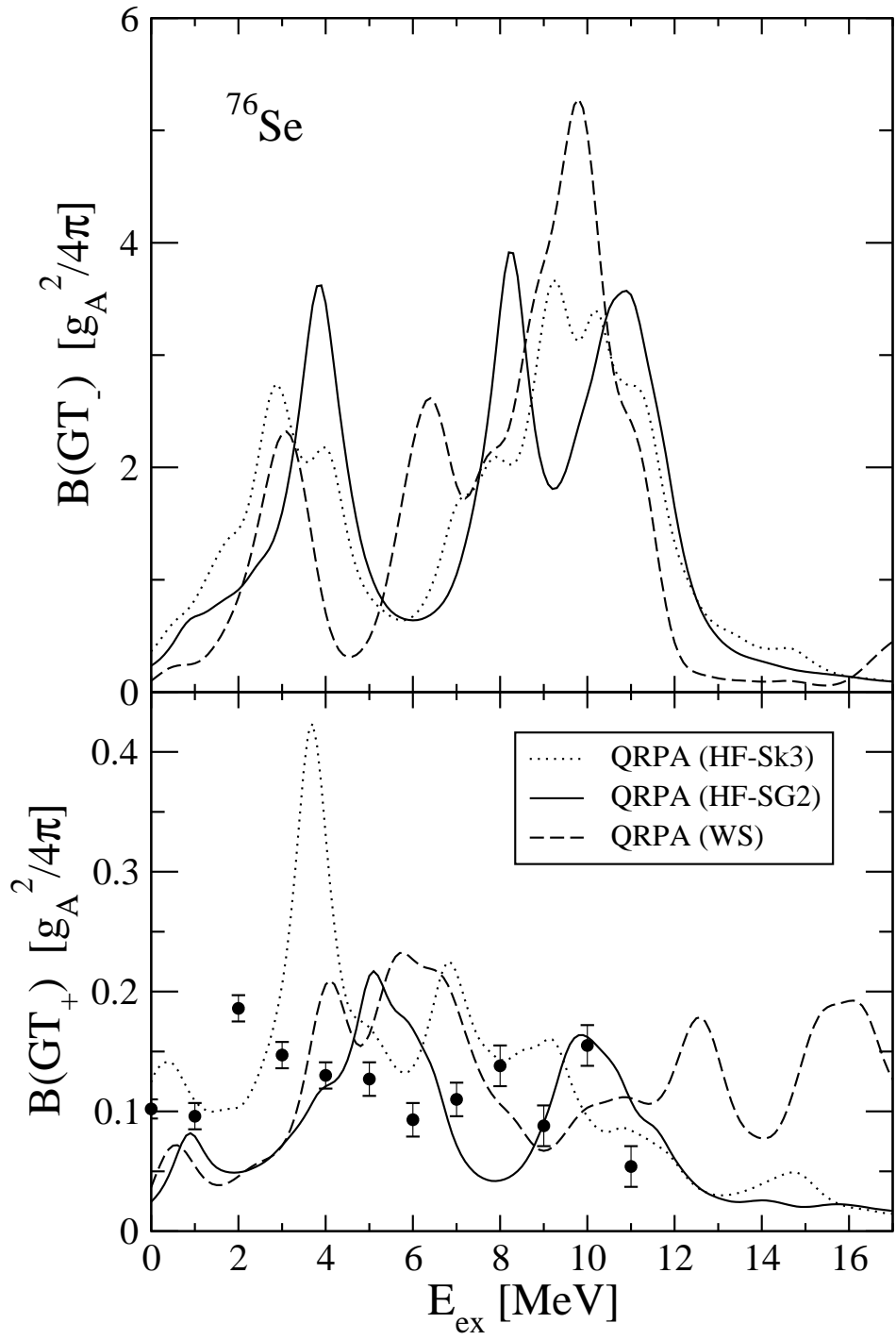}
\vskip 1cm
\caption{Same as in fig. 6 for $^{76}$Se. Data are from [29].}
\end{figure}

\begin{figure}[t]
\epsfig{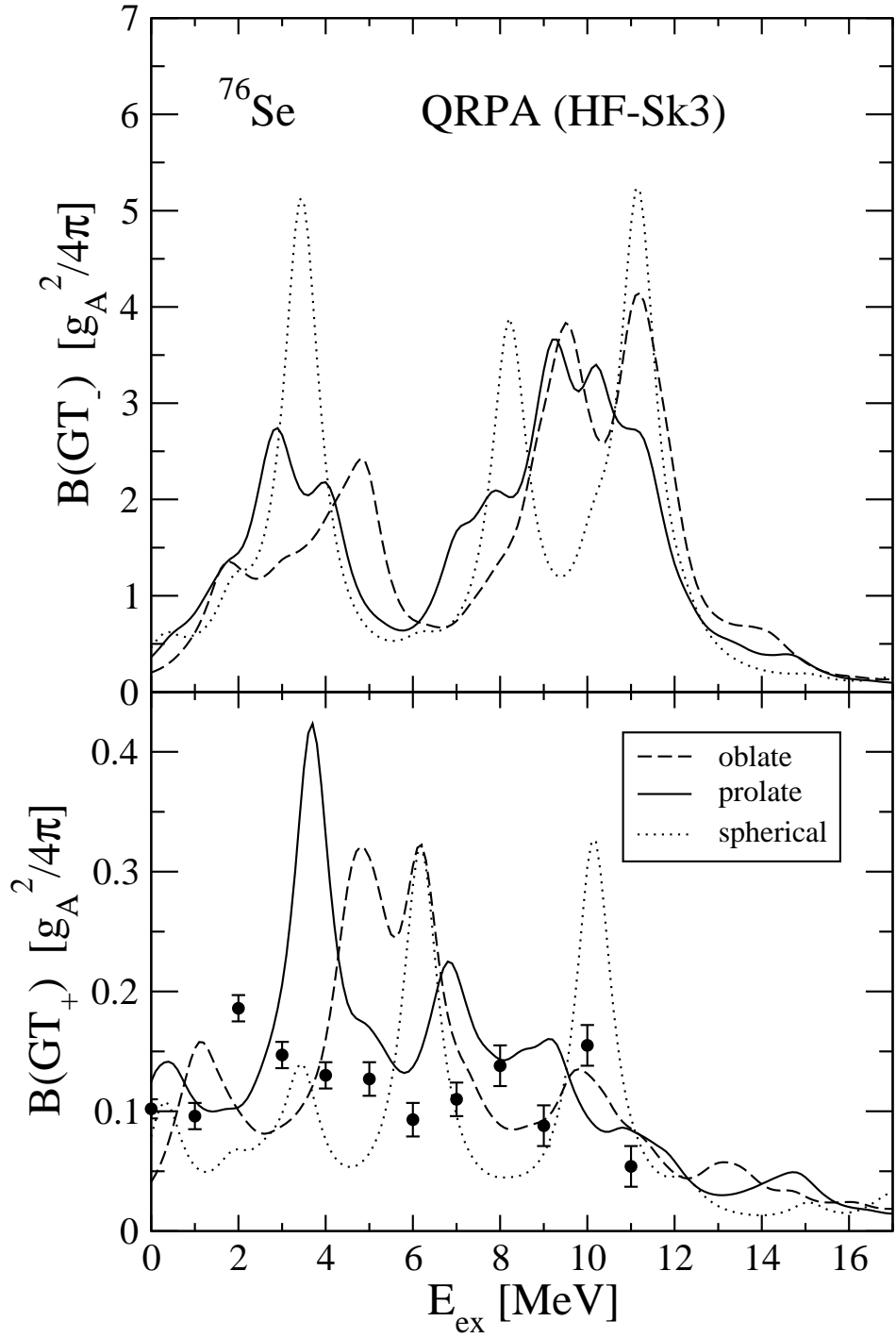}
\vskip 1cm
\caption{QRPA Gamow-Teller $B(GT_-)$ and $B(GT_+)$ strength distributions 
in $^{76}$Se. The calculations correspond to the force Sk3 for spherical
(dotted line), prolate (solid line) and oblate (dashed line) shapes. 
Data are from [29].}
\end{figure}

\end{center}

\end{document}